# Excess of $^{236}$U in the northwest Mediterranean Sea

E. Chamizo[a], M. López-Lora[a], M. Bressac[b,c], I. Levy[b], M.K. Pham[b]

[a] Centro Nacional de Aceleradores (Universidad de Sevilla, Consejo Superior de Investigaciones Científicas, Junta de Andalucía). Thomas Alva Edison 7, 41092, Seville, Spain.
[b] IAEA-Environment Laboratories, Monte Carlo, 98000, Monaco.
[c] Institute for Marine and Antarctic Studies, University of Tasmania, Hobart, TAS, Australia.

Abstract

In this work, we present first $^{236}$U results in the northwestern Mediterranean. $^{236}$U is studied in a seawater column sampled at DYFAMED (Dynamics of Atmospheric Fluxes in the Mediterranean Sea) station (Ligurian Sea, 43°25´N, 07°52´E). The obtained $^{236}$U/$^{238}$U atom ratios in the dissolved phase, ranging from about $2 \times 10^{-9}$ at 100 m depth to about $1.5 \times 10^{-9}$ at 2350 m depth, indicate that anthropogenic $^{236}$U dominates the whole water column. The corresponding deep-water column inventory (12.6 ng/m$^2$ or $32.1 \times 10^{12}$ atoms/m$^2$) exceeds by a factor of 2.5 the expected one for global fallout at similar latitudes (5 ng/m$^2$ or $13 \times 10^{12}$ atoms/m$^2$), evidencing the influence of local or regional $^{236}$U sources in the western Mediterranean basin. On the other hand, the input of $^{236}$U associated to Saharan dust outbreaks is evaluated. It is estimated an additional $^{236}$U annual deposition of about 0.2 pg/m$^2$ based on the study of atmospheric particles collected in Monaco during different Saharan dust intrusions. The obtained results in the corresponding suspended solids collected at DYFAMED station indicate that about 64% of that $^{236}$U stays in solution in seawater. Overall, this source accounts for about 0.1% of the $^{236}$U inventory excess observed at DYFAMED station. The influence of the so-called Chernobyl fallout and the radioactive effluents produced by the different nuclear installations allocated to the Mediterranean basin, might explain the inventory gap, however, further studies are necessary to come to a conclusion about its origin.

Keywords: $^{236}$U, seawater, western Mediterranean Sea, AMS, Saharan dust outbreaks

1. Introduction

The study of artificial radioactivity in the marine environment has contributed in no small way to ocean science. The introduction of huge amounts of radioactivity in the world oceans as a consequence of the atmospheric nuclear tests performed between 1945 and 1980 drove in the first two decades, roughly from 1950 to 1970, ocean studies at the national and institutional level with limited coordination. The RIME (Radioactivity in the Marine Environment) report provided an account of what had been learned about radionuclides in the oceans since 1957 [1]. In the 1970's, the GEOSECS (Geochemical Ocean Sections Study) project [2, 3] produced the first comprehensive data set for the fallout radionuclides $^{137}$Cs ($T_{1/2}$=30.07 y), $^{90}$Sr ($T_{1/2}$=29.79 y), $^{239,240}$Pu ($T_{1/2}$=24110 and 6524 y, respectively) and limited $^{241}$Am ($T_{1/2}$= 432.2 y) for the Atlantic and Pacific Oceans, providing first estimates of inventories and fallout patterns. It also compiled even more detailed data sets for $^{3}$H ($T_{1/2}$=12.33 y) and $^{14}$C ($T_{1/2}$=5730 y) in, besides, the Indian Ocean, prompting the use of these conservative radionuclides by the physical oceanographic community. Subsequently, the fallout radionuclides were classified in two categories according to their physic-chemical properties: the conservative or soluble radionuclides (e.g. $^{90}$Sr and $^{137}$Cs [4]), that would be moved according to the physical dispersive and mixing processes in the oceans; and the non-conservative or particle-reactive ones (e.g. plutonium and americium isotopes [5]), that would have patterns mainly driven by the presence of organic and inorganic particles, being more abundant in sediments. In the following 20 years, this preliminary knowledge was used mainly to evaluate the impact of new sources of radioactivity on local and regional scales. For instance, the discharges of radioactive wastes to the Irish Sea from the British nuclear fuel reprocessing plant (NRP) at Sellafield, and the ones to the English Channel from the French NRP at Cap de La Hague, impacted the oceans and seas around Europe. Indeed, the quantities of radioactive wastes discharges from the Sellafield NRP have been considered the largest intentional discharge so far made by the nuclear industry into the marine environment. Another illustrative example is the accident at the Chernobyl nuclear power station (NPS) in 1986 (former Soviet Union, now Ukraine), which released huge amounts of different radionuclides all over Europe, impacting the Baltic, Black and Mediterranean Seas and the European and Arctic shelves and enclosed seas. This accident was classified in 2006 by the International

Atomic Energy Agency (IAEA) as the "foremost nuclear catastrophe in human history", a perception that still remains valid even after the Fukushima accident [6, 7].

During the more than 60 years of the use of artificial radioactivity in ocean science, studies have been driven by the state of the art of the analytical techniques and sampling technologies, and their availability. The 1960´s and the 1970´s benefited from the advances in solid-state detectors and in the technology of the associated electronics. The arrival of surface barrier silicon detectors made it possible the use of long-lived alpha emitting radionuclides, including the important group of transuranic elements, to trace oceanic processes on long time scales [8, 9]. In the following 20 years, Mass Spectrometry (MS) techniques, where atoms are counted directly, have been used more frequently for plutonium analysis. They provide also information on $^{240}$Pu/$^{239}$Pu atom ratio, which informs about the plutonium source. In general, MS techniques have the potential to produce better results due to their significantly better overall counting efficiency compared to radiometric techniques. A direct consequence is the reduction of the sample size (e.g. from tens of liters down to 10 L of seawater) with the subsequent simplification of the sample processing, and of the measuring time (i.e. from days to less than 1 hour of counting) [10, 11].

To date, among the different available MS techniques, Accelerator Mass Spectrometry (AMS) and Inductively Coupled Plasma Mass Spectrometry (ICPMS) are the two most competitive techniques in the field of plutonium isotopes determinations [12, 13]. A large number of facilities worldwide have demonstrated their potential to measure them at environmental levels and many sets of results have come to light. As a consequence, much has been learned about the distribution and behavior of plutonium in the oceans [6].

The situation has been different for $^{237}$Np and $^{236}$U, two anthropogenic radionuclides with many potential applications in oceanography due to their conservative nature in seawater, but scarcely studied due to technical limitations of different nature. $^{237}$Np ($T_{1/2}$=2.14 My) is produced by $^{238}$U(n,2n)$^{237}$U→$^{237}$Np ($\sigma$=0.8 barn) reaction through fast neutrons in nuclear bomb testing and reactors, and by $^{235}$U(n,γ)$^{236}$U(n,γ)$^{237}$U→$^{237}$Np reaction through thermal neutrons in nuclear reactors. It can be also produced in lesser amounts from $^{241}$Pu ($T_{1/2}$=14.29 y) via the alpha-decay of its daughter, $^{241}$Am. Although it is in the scope of the former MS techniques, the complexity of the chemical procedures needed to purify the neptunium fraction with a reliable control of the yield, has limited its analysis on a routine basis [14]. Only a few results in very specific scenarios have been published (e.g. Baltic Sea, impacted by the Chernobyl accident [15]). $^{236}$U ($T_{1/2}$= 23.4 My) is produced mainly in two nuclear reactions that can occurred either naturally or be induced by man: by $^{235}$U(n,γ)$^{236}$U (σ=86.7 barn) reaction through thermal neutrons, and by $^{238}$U(n,3n)$^{236}$U (σ=0.4 barn) reaction when fast neutrons are involved. The third way of production is the growth from $^{240}$Pu deposited already on land by global fallout [16]. In this case, the limitations have been of technical nature. In seawater, $^{236}$U is present at isotopic abundances (i.e. $^{236}$U/$^{238}$U atom ratios) that range from about $10^{-13}$ (pre-nuclear levels [17]) to $10^{-6}$ (seawater from the Irish Sea impacted by the Sellafield NRP [18]). Therefore, MS techniques covering that range are necessary to perform general environmental studies. To date, AMS offers the lowest $^{236}$U/$^{238}$U background ratios in environmental matrixes, mainly thanks to the destruction of molecular isobars in the so-called stripping process. At AMS facilities with accelerator terminal voltages of 3 MV and above and with an optimized setup, ratios as low as $10^{-13}$ have been reported [19, 20]. Recently, two compact AMS facilities (i.e. terminal voltages of 1 MV and below) have demonstrated their potential to measure $^{236}$U at environmental levels: the 600 kV AMS system at the ETH Zürich (i.e. with a $^{236}$U/$^{238}$U background ratio at the level of $10^{-13}$ [21]), and the 1 MV AMS system at the Centro Nacional de Aceleradores (CNA, Seville, Spain), with a sensibility below $10^{-10}$ [22, 23].

The first feasibility study for using $^{236}$U as an oceanic circulation tracer was performed in [24]. It was reported first experimental evidences about its conservative nature in seawater based on a comparative study of the depth profiles of $^{137}$Cs and $^{236}$U in the Japan/East Sea. At the same time, first $^{236}$U results on two oceanic depth profiles sampled in the western equatorial Atlantic Ocean were reported [17]. The detection of anthropogenic $^{236}$U at abyssal depths suggested the influence of the liquid effluents released by the European NRP, through the so-called North Atlantic Deep Water formation. Later on, this result was confirmed through the study of a transect along the Northwest Atlantic Ocean [25]. On the other hand, it has been demonstrated that the $^{129}$I/$^{236}$U ratio can be used as a new (quantitative) tracer for water mass transport times in the Arctic Ocean and beyond [21]. Also in the scope of oceanography, it has been estimated the input of $^{236}$U into the ocean over the past decades through the study of a coral record from the Caribbean Sea [26]; and it has been studied the $^{236}$U/$^{238}$U atom ratio in marine sediments affected by different local sources [27]. These first results have attracted attention to the fact that, still, much more research is needed to fully understand the sources of $^{236}$U to the marine environment and its biogeochemical behavior. This knowledge is necessary to make full use of the potential of $^{236}$U as an oceanic tracer.

In this work, we present first $^{236}$U results in the western Mediterranean. This area is interesting due to the potential local and regional $^{236}$U sources that might have impacted its waters and sediments, and due to the natural processes that might have altered the fallout inventories for the different radionuclides [28]. The northwest Mediterranean Sea washes the shores of France, which is the largest consumer of nuclear energy in the world. One-third of the NPS´s have been allocated to the Mediterranean basin, together with the second most important NRP in France (Marcoule). Therefore, that area might be impacted by additional $^{236}$U sources other than global fallout: the liquid effluents released by Marcoule NRP, and the wastes derived from the normal operation of the different NPS´s. On the other hand, the Mediterranean region has been subjected to the relatively strong influence of the Chernobyl fallout. First measurements suggested that most of the refractory elements were deposited close to the stricken reactor (i.e. 100 km around the NPS) [29]. However, hot particles have been found in at least 15 European countries far away from the NPS [30], and $^{236}$U results in soils at distances more than 200 km to the north of the plant might indicate the impact of Chernobyl fallout [31]. These results could evidence a more widespread dispersion of this radionuclide [31]. Here, the qualitative importance of these sources is discussed through the study of $^{236}$U in a seawater column from the so-called DYFAMED (Dynamics of Atmospheric Fluxes in the Mediterranean Sea) station, in the western Mediterranean. On the other hand, the western Mediterranean region is seasonally impacted by the entrance of soil particles coming from the North of Africa (mainly Marocco and Algeria). In the 1960´s, France conducted in Reggane (central Algeria) a set of four nuclear tests code-named "Gerboise" with a total released yield between 40 and 110 ktons TNT-equivalent. IAEA surveys concluded that the radioactivity associated to the sand fraction was very low, being most it concentrated at ground-zero and associated to fused sand aggregates, which are too heavy to be carried by wind [32]. Therefore, crustal particles associated to Saharan dust outbreaks contain mostly artificial radionuclides coming from global fallout. As it is the case of plutonium isotopes [33], $^{236}$U associated to those events might represent an important fraction of the current annual atmospheric input to the Mediterranean region. In order to find out about this additional input, $^{236}$U has been studied in sediment trap samples collected also at DYFAMED station and in dust particles collected in Monaco during different Saharan Dust intrusions [34]. Next, the details of the sample preparation procedure, the measurement technique, the results and the main conclusions are presented.

2. Materials and methods

2.1. Samples

The Mediterranean Sea is a semi-enclosed deep basin connected to the Atlantic Ocean through the Strait of Gibraltar. It can be subdivided in two basins, the western and the eastern ones, through the Strait of Sicily. For this study, seawater samples collected at DYFAMED station in the Ligurian Sea, on the western basin, were taken (43°25´N, 07°53´E, Fig. 1). That station has been the base of numerous short and long-term research programs in the last 25 years, and it is considered one of the best studied locations of the world's oceans, and a reference to monitor ongoing changes in the Mediterranean environment [35]. The samples were collected at 5 different depths (surface, 100 m, 300 m, 1000 m and 2000 m depth) covering the whole seawater profile (2350 bottom depth). 30 L Niskin bottles mounted on a Rosette sampler were employed for the sampling [33]. Approximately 60 L of unfiltered seawater were taken at each depth and acidified with HCl to pH 1-2.

Saharan dust samples were collected on the roof of the IAEA Environment Laboratories (IAEA-EL) premises in Monaco (43°45´N, 07°25´E) during four episodic Saharan dust events: the one detected on the night of the 20[th] of February 2004, where significant quantities of red-colored particles were deposited; an another three less important events (1[st] of May 2004, 29[th] of July 2005 and 26[th] of May 2008). Atmospheric input into the water column during the 2004 events is also studied using sediment trap samples collected at 200 m and 1000 m depth at DYFAMED station. The elemental composition and the presence of natural and anthropogenic radionuclides in those atmospheric samples have been reported in [36].

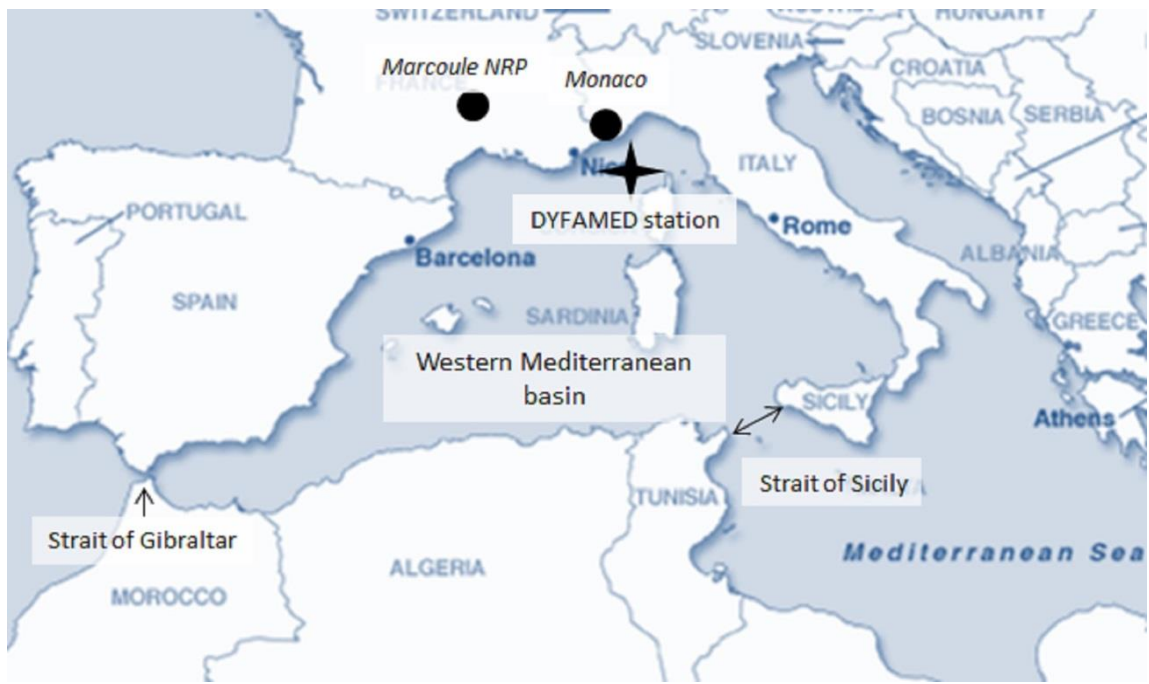

Figure 1.- Map of the western Mediterranean basin with the situation of the most relevant locations: the DYFAMED observatory (43°25´N, 07°53´E); the city-state of Monaco; Marcoule NRP, 120 km upstream the Rhone River mouth; and the two Straits limiting the basin, Strait of Gibraltar and Strait of Sicily.

2.2. Radiochemical procedure

2.2.1. Reagents and spikes

For $^{236}$U AMS determinations, it is very important to minimize $^{238}$U and $^{236}$U contaminations introduced by reagents, spikes and laboratory materials. This is especially critical when small samples are processed (e.g. 1 g of sediment or 5 L of seawater, for instance), as $^{238}$U and $^{236}$U contaminations can bias the obtained $^{236}$U/$^{238}$U and $^{236}$U/$^{233}$U atom ratios, which are the results of an AMS measurement. To prepare the AMS samples in this work, glassware was avoided and replaced by Teflon® when possible; an ultra-pure $Fe^{3+}$ standard solution provided by High Purity Standards (HPS, England), verified for the absence of uranium isotopes, was used for the $Fe(OH)_3$ precipitation step during the cathode preparation; and acids of the highest purity were used for the final purification of the uranium fraction. In order to control the $^{236}$U and $^{238}$U backgrounds that might have been introduced at IAEA-EL premises in Monaco, where no special protocols to avoid uranium contamination are adopted given the large volumes of seawater that are routinely processed, two procedural blanks (i.e. distilled water processed in the same way as DYFAMED seawater samples) were also prepared and measured by AMS. It was obtained $^{238}$U currents and $^{236}$U count rates at least 70 times and 30 times lower, respectively, than the ones got from the samples of interest. On the other hand, the presence of $^{236}$U in the $^{233}$U and $^{242}$Pu spikes was also verified by measuring a set of unprocessed spiked AMS blanks (i.e. $^{233}$U or $^{242}$Pu dispersed in an iron oxide matrix). $^{233}$U standard solution was provided by IAEA-EL (unknown supplier); $^{242}$Pu by National Physical Laboratory (NPL, United Kingdom). $^{236}$U was only detected in the $^{233}$U tracer, with an atomic abundance of $5 \times 10^{-4}$. This result is slightly better than the one reported in a previous study [27], probably because of the use of a cleaner equipment. The samples of interest produced $^{236}$U/$^{233}$U atom ratios at least two orders of magnitude higher. Therefore, the corresponding background corrections are in every case negligible.

2.2.2. Description of the radiochemical procedures

At the time the analysis of $^{236}$U was considered, the samples had been already processed at the IAEA-EL laboratories in Monaco for the separation and purification of cesium, strontium, plutonium and neptunium fractions. Therefore, uranium was recovered following a different approach: the residues produced throughout the former procedure containing most of the uranium were further processed. Details about the final method are given in Fig. 2.

The chemical procedure performed in a first instance at IAEA-EL premises in Monaco, consisted of three main stages: (i) separation of actinides from the bulk sample using two sequential $MnO_2$ and $Fe(OH)_2$ precipitation steps; (ii) removal of major matrix components of the sample and a first separation of Np and Pu fractions using an AG-1X8® resin; and (iii) a second purification of those fractions using additional chromatography resins [37]. The AG-1X8® resin does not retain uranium when the sample is introduced in 8M $HNO_3$ as it is the case, so the corresponding washings were combined and taken as the starting point for this study. This secondary sample (from now on, "AMS sample") was further processed: it was evaporated to dryness several times and, once dissolved in 3M $HNO_3$, uranium was separated by $Fe(OH)_3$ precipitation by adjusting the pH to 8-9 with $NH_4$. As the AMS sample contained most of the matrix components of the original sample and added reagents, no additional iron was necessary for this third concentration step. Once dried, the $Fe(OH)_3$ samples were taken to the CNA laboratories in Seville, Spain, where uranium was finally separated as it follows: the samples were dissolved in 3M $HNO_3$; spiked with 3 pg ($10^{-12}$ g) of $^{233}U$; and introduced into UTEVA® resins for the final uranium purification, following the guidelines given in [38]. For the AMS cathode preparation, 1 mg of $Fe^{3+}$ was added to the corresponding uranium solutions to precipitate it with $Fe(OH)_3$. These precipitates were transferred to quartz crucibles, dried, baked at 650 °C for 1 h to convert the uranium and the iron to the oxide form, and finally mixed with about 3 mg of Nb powder and pressed into appropriate aluminum cathodes.

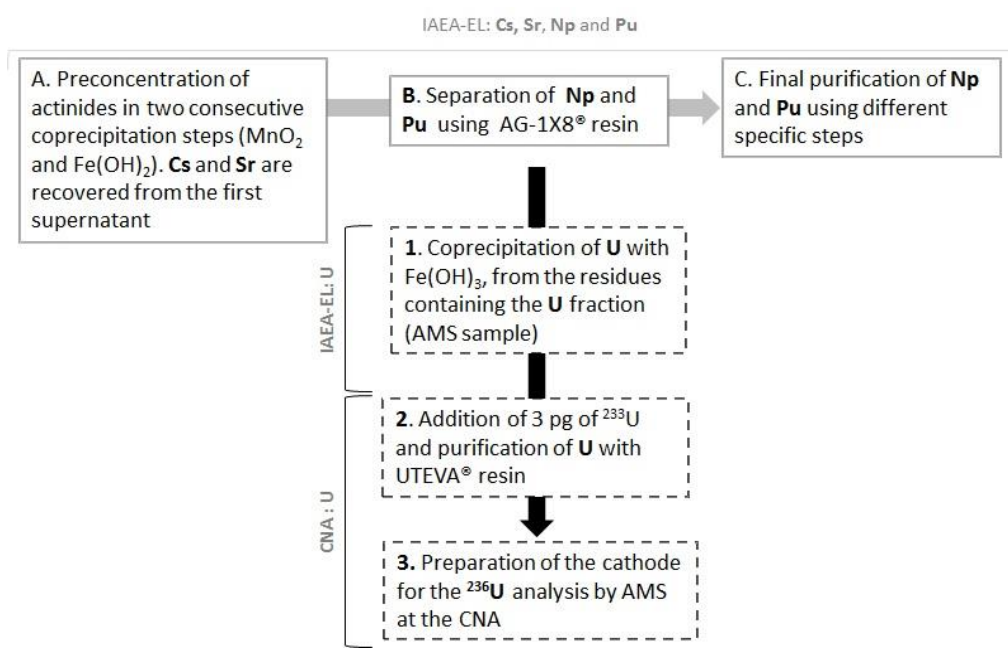

Figure 2.- Diagram showing the radiochemical procedure followed in this work to separate uranium from DYFAMED seawater samples. The first steps of the procedure were carried out at the IAEA-EL laboratories in Monaco (steps A, B, C and 1). Uranium was finally purified at CNA (steps 2 and 3). Details about the IAEA-EL procedure for the sequential separation of the Cs, Sr, Np and Pu fractions can be found in [9].

Saharan dust and sediment trap samples were fully processed at CNA laboratories for the sequential separation of plutonium and uranium fractions, following a similar procedure as the one given in [27] for solid samples with no heterogeneities. 1 g aliquots were spiked with about 3 pg of $^{242}$Pu and $^{233}$U, and thoroughly homogenized with a few ml of 0.1M HCl. Once dried, the samples were calcined at 600ºC in a muffle furnace, and the resulting ashes were leached in a hot plate with 8M $HNO_3$ and $H_2O_2$ for about 5 hours at 90ºC. That way, anthropogenic radionuclides attached to the surface of the crustal components of the samples were put in solution. Actinides were then concentrated from the resulting supernatant with $Fe(OH)_3$. Finally, uranium and plutonium were separated and purified from that iron precipitate using TEVA® and UTEVA® resins following the guidelines given in [38, 39]. The U and Pu AMS cathodes were produced from the final solutions as described before, and the analyses were performed also on the 1 MV AMS facility at the CNA in 2014. $^{239,240}$Pu results have been reported in [36].

2.3. AMS measurement of $^{236}$U

Uranium isotopes ($^{233}$U, $^{236}$U and $^{238}$U) were measured on the 1 MV AMS facility at CNA in 2014. Details about the technique can be found in [22]. Briefly, uranium isotopes are extracted as oxide anions ($^{233,236,238}U^{16}O^-$) in a Cs-sputter ion source; analyzed by a 90º sector magnet; injected into the tandem accelerator working at about 700 kV where they are stripped to 3+ ($^{x}U^{3+}$); and further analyzed by a 90º sector magnet and a 120º electrostatic deflector. During these measurements, Ar gas was available as stripper, giving an overall $U^{3+}$ transmission through the accelerator of about 11%. The minor isotopes ($^{236}$U and $^{233}$U) are counted from the total energy signal provided by a gas ionization chamber, placed at the end of the beam line. $^{238}U^{3+}$ is measured as a beam-current in an off-axis Faraday cup placed at the exit of the 90º magnet on the high-energy side (FC3). During a routine measurement, the cycling between the different masses is performed using the so-called Slow Sequential Injection System, which allows the sequential change, in second pulses, of different parameters of the AMS system. In our case, $^{238}$U is guided to a certain FC3 position by adjusting the voltage of the chamber of the low-energy magnet (the so-called bouncer) and the terminal voltage; for the minor radionuclides, the 120º electrostatic deflector is also modified, in order to guide the particles into the gas detector. Every sample is usually measured 11 times or runs together with standards, instrumental and procedural blanks.

Instrumental blanks are not processed in the laboratory, so they provide information about the contamination introduced during the AMS measurement. In every run, $^{238}$U, $^{236}$U and $^{233}$U are measured 5 times in pulses of 5, 20 and 3 seconds, respectively. Overall, about 25 min of total counting time is dedicated to every sample. The instrumental precision of the measurement is 2%.

The final AMS results are $^{236}$U/$^{238}$U, $^{233}$U/$^{238}$U and $^{233}$U/$^{236}$U atom ratios. These ratios have to be corrected by the instrumental and procedural blanks, and by the $^{236}$U background associated to the $^{233}$U spike. Once corrected and assuming an identical chemical behavior for every isotope, the obtained $^{236}$U/$^{238}$U atom ratio characterizes the sample of interest, and the $^{236}$U/$^{233}$U and $^{236}$U/$^{233}$U ratios can be used to quantify the $^{236}$U and $^{238}$U concentrations in the original sample based on the so-called isotope-dilution method. For solid samples (i.e. Saharan dust and sediment trap samples), it is important to remark that, in our case, the so-obtained $^{238}$U concentrations are not representative of the original samples, as only a fraction of the natural uranium is put in solution when applying a conventional leaching method [16]. This is not the case of seawater where the originally dissolved uranium is chemically separated. However, in the case of the studied DYFAMED seawater samples, the $^{236,238}$U concentrations couldn´t be obtained from those experimental figures, as they were spiked with $^{233}$U in an intermediate stage of the procedure (Fig. 2). In order to find out about those concentrations in the original seawater samples, an alternative approach was considered. The empirical formulas that relate the $^{238}$U concentrations to salinity in seawater were used [40, 41], and $^{236}$U concentrations were calculated based on the measured $^{236}$U/$^{238}$U AMS ratios (Table II). The resulting $^{238}$U concentrations in the so-called AMS samples were used to get further information about the chemical yield of the procedure. By comparing them with the calculated ones from the salinity data, the uranium yield for the stages of the procedure carried out at the IAEA-EL was estimated. The conclusion is that about 48% of the uranium present in the original 60 L seawater aliquots was finally processed for the $^{236}$U AMS determinations (Table I). During the analysis of the DYFAMED seawater samples, $^{238}$U$^{3+}$ beam-currents of about 0.4 nA were monitored in FC3. This means that it was obtained about 4 pA of $^{238}$U$^{3+}$ beam-current per µg of $^{238}$U in the sample given the previous yield or, equivalently, 8 cps in the detector per pg of uranium for the minor isotopes ($^{236}$U or $^{233}$U). Therefore, the corresponding counting efficiency for $^{236}$U was at the level of 3x10$^{-6}$ during those experiments.

3. Results

3.1. $^{236}$U in the western Mediterranean- results for seawater samples

In Table II and Fig. 3, it is displayed the obtained $^{236}$U results ($^{236}$U concentrations and $^{236}$U/$^{238}$U atom ratios) for the studied DYFAMED seawater samples. As it can be observed, $^{236}$U/$^{238}$U atom ratios show a decreasing trend with depth and are all between $1.5 \times 10^{-9}$ and $2.1 \times 10^{-9}$. They are at least four orders of magnitude higher than the natural oceanic signal, probably at the level of $10^{-13}$ [17]. Therefore, it can be concluded that DYFAMED seawater column is dominated by anthropogenic $^{236}$U. The surface $^{236}$U/$^{238}$U ratio, of about $2.1 \times 10^{-9}$, exceeds by a factor of two the estimated one for modern ocean surface waters, of $1 \times 10^{-9}$ [17], and by about 45% the one measured in surface waters from the Japan Sea at a similar latitude (40º25.66'N), of about $1.5 \times 10^{-9}$ [24]. The deepest sample (2350 m depth) shows also a ratio at the level of $10^{-9}$, 1.4 times lower than the surface one. The estimated $^{238}$U mass concentrations in the water column based on salinity data are about $3.5 \pm 0.2$ µg/L (Tables I and II). It is interesting to mention that $^{236}$U shows the same trend along the profile as $^{137}$Cs and $^{237}$Np, another conservative radionuclides that were also measured from the same seawater aliquots, whose results will be reported elsewhere [42]. Indeed, the obtained $^{237}$Np/$^{236}$U and $^{137}$Cs/$^{236}$U atom ratios for the individual samples stay constant, with average and standard deviation values of $1.37 \pm 0.09$ and $0.123 \pm 0.003$, respectively. The $^{236}$U atom concentrations range from about $13 \times 10^6$ to $18 \times 10^6$ atoms/L (Table II and Fig. 3). They are remarkably higher than the measured ones in the western equatorial Atlantic Ocean, from about $1 \times 10^6$ to $6 \times 10^6$ atoms/kg [17] and, at least, 30% higher the ones obtained for the Japan Sea, from about $1.5 \times 10^6$ to $13 \times 10^6$ atoms/kg [24]. The corresponding $^{236}$U deep water-column inventory (0-2350 m) at the studied site is $32.1 \times 10^{12}$ atoms/m$^2$. It exceeds by more than factor of 2 the estimated one for global fallout at similar latitudes, as it will be discussed in section 4.1. In contrast to this, the obtained one for $^{237}$Np is in agreement with the expected one for that source ($23 \times 10^{12}$ atoms/L) [42]. These results might indicate the impact of local or regional sources of $^{236}$U to the western Mediterranean Sea.

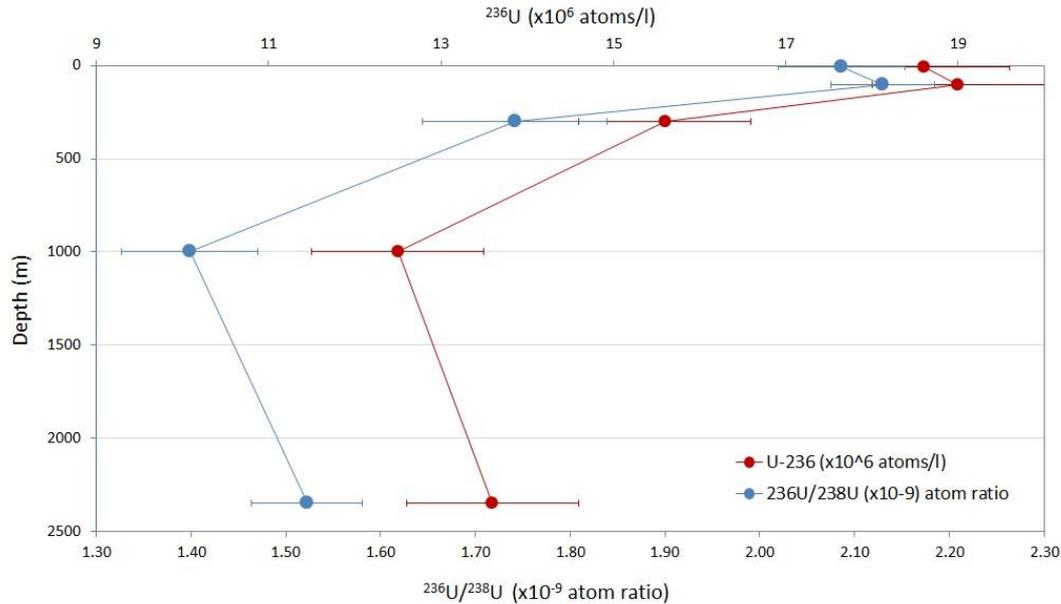

Figure 3.- Depth profiles of $^{236}$U concentrations (in units of $10^6$ atoms/L, red legends) and $^{236}$U/$^{238}$U atom ratios (in units of $10^{-9}$, blue legends) at the DYFAMED station. Numeric results displayed in Table II.

3.2.$^{236}$U in the western Mediterranean- results for Saharan dust samples and sediment traps

In Table III, it is displayed the obtained $^{236}$U results in atmospheric particles collected in Monaco during the four studied Saharan dust intrusions. In Table IV, it is presented the obtained results in sediment trap samples collected at DYFAMED station during the 20$^{th}$ of February and 1$^{st}$ of May 2004 events. In every case, $^{236}$U/$^{238}$U atom ratios were calculated considering the $^{236}$U AMS concentrations and the $^{238}$U ICP-MS data reported in [36], which were obtained following a total decomposition of the samples. Also in Tables III and IV it is displayed the $^{236}$U/$^{239}$Pu atom ratios, which were calculated using the $^{239}$Pu AMS results obtained in the same sample aliquots and also reported in [36].

The estimated $^{236}$U/$^{238}$U atoms ratios for the Saharan dust samples collected in Monaco, of about $10^{-8}$, are in reasonable agreement with the expected ones for soils solely influenced by global fallout [16]. The same result was obtained from the study of the $^{240}$Pu/$^{239}$Pu atom ratio in the same samples in a previous work, with an average value of 0.20±0.01 [36], in accordance with the expected one for global fallout at those latitudes (0.178±0.019 [43]). Thus, the observed ratios are consistent with the most likely origin of those atmospheric particles: resuspension of soils from Northern Africa containing global fallout radionuclides. It is interesting to remark that the $^{236}$U concentrations and the $^{236}$U/$^{239}$Pu atom ratios in the deposited dust were very similar in the four studied events (Table III), which might indicate a similar emission source. When entering the water column, the uranium isotopic composition of those particles changes significantly. The obtained results in sediment trap samples during the two sampling periods in 2004 evidence that (Table IV). Whereas the $^{238}$U concentration decreases by (23±2)% (from 2.2±0.2 µg/g in dust particles to 1.7±0.06 µg/g in sediment trap samples), the $^{236}$U one diminishes on average by (63±4)% (from (57.4±4.5)x10$^6$ atoms/g to (21±7)x10$^6$ atoms/g) (Tables III and IV). This might indicate a higher solubility in seawater of anthropogenic $^{236}$U (which is attached to the surface of soil particles) compared to natural $^{238}$U (which is immersed in the crustal lattice [16]). Therefore, the $^{236}$U/$^{238}$U atom ratios diminishes on average by a factor of about 2 (from (10.5±1.4)x10$^{-9}$ to (5±1.6)x10$^{-9}$, Tables III and IV). The measured $^{236}$U/$^{239}$Pu atom ratios show a similar effect due to the particle-reactive nature of plutonium, being a factor of about 3 lower in the suspended solids (from 0.11±0.02 to 0.031±0.016, Tables III and IV). It is interesting to remark that the $^{236}$U concentration in the sediment trap samples increased during the sampling period corresponding to the 20$^{th}$ of February 2004 event. That Saharan dust outbreak impacted significantly the Southern region of France, and has been studied by other authors [34].

4. Discussion

4.1. Sources and inputs of anthropogenic $^{236}$U into the western Mediterranean basin

As it was previously stated, the so-called general fallout represents the most significant source of $^{236}$U to the general environment. The estimated global inventories range from 900 kg (based on soils from Japan, [16]) to 1060 kg (based on a coral core from Caribbean Sea, [26]). However, any systematic study on the latitudinal distribution of $^{236}$U has been carried out so far. At the studied site, there is no experimental information on the $^{236}$U global fallout inventory. However, it can be estimated using two parameters: (i) the $^{236}$U/$^{239}$Pu ratio measured in soils free of local or regional sources close to the area of interest, and (ii) the reported $^{239}$Pu inventories at similar latitudes. Recently, it has been published a value for the $^{236}$U/$^{239}$Pu atom ratio of 0.23±0.021 for a reference soil sample from Austria (IAEA-Soil-6) collected in the pre-Chernobyl period [22]. This value is in agreement with the reported one for soils from Japan, ranging from 0.21 to 0.25 [16]. Both results point out to a homogenous distribution of $^{236}$U/$^{239}$Pu global fallout atom ratio in the northern latitudes, making its use reasonable for further estimations. On the other hand, it is assumed that the $^{239}$Pu fallout inventory at the studied site is the same as the one reported for Monaco, a city-state located at the Mediterranean coast 54 km away, of 49±4 Bq/m$^2$ ((5.4± 0.4)x10$^{13}$ atoms/m$^2$ or (21±2) ng/m$^2$ [28]) (Fig. 1). From those two figures, a $^{236}$U global fallout inventory at the DYFAMED site of 5±1 ng/m$^2$, or (1.3±0.2)x10$^{13}$ atoms/m$^2$, can be estimated. This is consistent with the obtained deep water-column inventory at the Japan Sea, of (1.37±0.09)x10$^{13}$ atoms/m$^2$ [24], however, it is a factor of 2.5 lower than the one obtained at the studied site, of (3.2±0.2)x10$^{13}$ atoms/m$^2$ or 12.6±0.8 ng/m$^2$.

It is reasonable to consider the obtained $^{236}$U results at DYFAMED station representative of the whole western basin as an estimate study. Two main reasons support this assumption. (i) The site is generally accepted as an open-ocean site, where the atmosphere constitutes the only significant source of natural lithogenic and anthropogenic material. It is separated from coastal lateral inputs by a strong horizontal density gradient caused by the permanent geostrophic Ligurian frontal jet flow, and is globally isolated from riverine inputs by the Northern Current. Therefore, it is not directly influenced by terrestrial local sources [35]. (ii) It is likely that anthropogenic radionuclides that behave conservatively in seawater have achieved almost steady-state in the western basin. This assumption is based on the fact that no changes in the $^{237}$Np inventories and vertical distributions have been observed at DYFAMED station between 2001 and 2013 in a recent publication [42]. In concentration basins as it is the case of the Mediterranean, evaporation increases the salinity of the surface waters, raising their density and producing convection. Deep water renewal is therefore a nearly continuous process, and the waters of the basin are well ventilated at all depths. Therefore, it is expected that anthropogenic radionuclides are quickly distributed in the western basin. Indeed, $^{137}$Cs was found at 2000 m depth at DYFAMED station in 1976, 13 years after peak of the atmospheric nuclear testing [44]. Thus, the obtained $^{236}$U inventories can be integrated over the surface of the whole western Mediterranean basin ($8.5 \times 10^5$ km$^2$ [33]) as a first approximation, and insights of the different input sources to the whole basin can be worked out.

The result is that about 10.7 kg of $^{236}$U would have been distributed in that area, 4.25 kg of them would have been introduced during the period of atmospheric nuclear testing. It is important to remark that such a difference hasn´t been obtained for the other actinides studied at the same site, such as $^{237}$Np and $^{239,240}$Pu [42]. The $^{236}$U inventory gap, of about 6 kg, might reflect the impact of local or regional $^{236}$U sources, never investigated before. On the other hand, additional $^{236}$U contributions can be attributed to well-known natural processes affecting the western Mediterranean, such as the exchange of seawater via Straits, the effect of rivers run-off, or the deposition of particles associated to Saharan Dust events. The different $^{236}$U additional inputs are summarized in Table V and are discussed next.

4.2. Natural processes affecting the $^{236}$U inventories in the western Mediterranean basin

It is well-known that there is a net in-flow of seawater from the Atlantic Ocean into the western Mediterranean Sea through the Strait of Gibraltar, and a net out-flow from the western into the eastern basins through the Strait of Sicily. The incoming waters, with a lower salinity, flow in the upper layers (from 200 to 400 m depth), which is more or less compensated by the outflowing saltier and denser waters in the deeper layers [28]. As a concentration gradient of $^{236}$U exists between both layers (Table II and Fig. 3), there would be a net flux that influences the water-column inventories. In the case of $^{137}$Cs, a net annual inflow from the Atlantic Ocean of 0.12 PBq/y and a net outflow to the eastern basin of 0.03 PBq/y was quantified up to 1986, increasing the global fallout inventory by 15% [28, 33]. As both radionuclides, $^{236}$U and $^{137}$Cs, show a similar depth profile due to their conservative nature in seawater [42], it is reasonable to assume a similar effect for $^{236}$U. This would mean an excess of about 0.6 kg of $^{236}$U in the whole western Mediterranean, or an additional inventory of about 0.7 ng/m$^2$, as it is shown in Table V.

Saharan dust events in the western Mediterranean have been recognized as important pathways of aerosol deliveries into surface waters. The Mediterranean is bordered on its southern and eastern shores by arid and desert regions, including the Sahara and Middle Eastern deserts. They act as sources of crust-dominated aerosols that are transported largely in the form of non-continuous dust pulses. The annual dust flux is controlled by a few of these episodes, accounting a single outbreak sometimes for more than 40% of the annual flux [45]. From the data in Tables III and IV, it is possible to get some information about the impact of these events on the $^{236}$U inventory at the studied site. Considering that the average annual dust deposition associated to these events in the 39°-42° northern latitude range in the western Mediterranean is of 10 g/m$^2$ [45], and that the measured $^{236}$U concentration in these dust samples is at the level of 0.02 pg/g (or about 6x10$^7$ atoms/g, Table III), an annual input of about 0.2 pg/m$^2$ is obtained. If this flux is assumed to be representative of the period 1963-2013 (i.e. from the year of maximum fallout to the sampling date of the studied DYFAMED seawater samples), a total $^{236}$U deposition of about 12 pg/m$^2$ is obtained, or about 8 g if integrated over the whole western basin. Taking into account that about 64% of that $^{236}$U is dissolved when entering the water column, this would add about 8 pg/m$^2$, or about 7 g if integrated, to the baseline $^{236}$U levels in the water column. This input is orders of magnitude lower than the other potential sources, as it was expected (Table V).

Finally, one has to consider the input of $^{236}$U associated to rivers run-off. Rivers are sources of freshwater and continental matter to the sea, carrying fallout radionuclides (if no other sources exist) weathered from the catchment basin or in solution. If soluble as it is the case of uranium isotopes, they would be transported along the river and enter the marine environment. To date, the published information on $^{236}$U in freshwater systems is very scarce. It has been reported an $^{236}$U atomic concentration of (2.5±0.7)x10$^7$ atoms/L in the Danube river, which flows into the Black Sea, and of (5.2±0.7)x10$^6$ atoms/L in Rio Negro, an effluent of the Amazonas river [46]. If it is assumed as a first approach that the Danube data represents the rivers washing the shores of the western Mediterranean basin, and it is considered the average freshwater fluxes to that basin for the Alboran, Southwestern, Northwestern and Tyrrhenian drainage basins (i.e. 5, 8, 91 and 18 km$^3$/y, respectively [47]), it can be concluded that about 60 g of $^{236}$U may have entered the studied area through rivers run-off from 1963 to 2013. Of course, this estimation is valid for indicating an order of magnitude, due to the uncertainty of the assumptions adopted. In any case, however, the $^{236}$U input associated to uncontaminated rivers represents, once more, a minor contribution compared to global fallout, as it is indicated in Table V.

## 4.3. Possible local and regional sources of $^{236}$U into the western Mediterranean

The $^{236}$U input associated to rivers run-off might be significant when contaminated regions are considered. This is the case of the Rhone River, which from 1961 to 2000 received the liquid effluents of Marcoule NRP, located 120 km upstream its mouth. To date, the released quantities of minor long-lived radionuclides haven´t been made public. However, the reported information for the two largest industrial NRP in Europe, Sellafield (UK) and La Hague (France), can be used to evaluate qualitatively the potential input of $^{236}$U from that nuclear facility. It has been estimated that about 90 kg of $^{236}$U were released by the Sellafield NRP, the largest one in Europe, into the Irish Sea from 1952, and about 25 kg by La Hague NRP into the English Channel from 1966 [25]. These amounts are not directly related to the capacities of the NRP´s, which are very similar in these two cases (i.e. of 1500 and 1700 t/y, respectively), but to other factors such as the type of irradiated fuel to be processed or the methodology of waste treatment adopted, which are not well documented [48]. Indeed, most of the $^{236}$U released by La Hague NRP and, possibly, by Sellafield NRP, was produced from 1980, presumably due to increasing utilization of reprocessed fuel in nuclear reactors [25]. In contrast to this, the maximum emissions of Pu and Np isotopes from both plants occurred in the 1970´s, being two orders of magnitude lower in the case of La Hague NRP [5, 28]. In the 1980´s, a similar rise was documented for Pu in the liquid effluents from Marcoule NRP. Indeed, traces of this Pu have been found in the sediments of the Rhone river estuary [49]. Taking into account all these factors and the fact that uranium is very soluble, it seems reasonable to consider an additional input of $^{236}$U from Marcoule NRP into the western Mediterranean, which might explain a significant fraction of the observed inventory gap. Further studies are necessary to confirm or refute this hypothesis.

Nuclear power stations allocated to the western Mediterranean basin in France, Italy and Spain, most of them along rivers, should be also considered as potential $^{236}$U sources. Again, there are difficulties for estimating the released quantities without control data, as the discharge rates depend on many factors such as the size and type of power reactor, the conditions of the reactor operation, and procedures used for waste treatment, among others [28]. However, there are experimental data indicating that a NPS can release $^{236}$U under routine operational conditions or during dismantling processes. A case study is the Garigliano NPS, located along the Garigliano river, in Italy [50]. Significantly higher $^{236}$U/$^{238}$U atom ratios have been found in sediments from the drain channel of the power station and 1 m downstream its mouth, evidencing the presence of irradiated uranium in the associated liquid effluents. However, no traces of that uranium were found upstream and 1 km downstream its mouth. Therefore, it is unlikely that the Garigliano NPS is responsible for the observed increase of $^{236}$U in the western Mediterranean. In any case, more experimental data is necessary in order to assess the $^{236}$U inventories coming from all these NPS´s.

The last potential $^{236}$U source to discuss is the so-called Chernobyl fallout. On the 26$^{th}$ of April 1986, the operational accident of the reactor No. 4 at the Chernobyl NPS located in former Soviet Union, now Ukraine, was destroyed as a result of a thermal explosion, causing severe radioactive fallout deposition over Europe beyond national boundaries. It has been estimated that about 13 kg of $^{239}$Pu (0.03 PBq) were released as particulate form at the time of the accident, accounting for about 2% of the cumulative amount of Pu in the reactor. The same percentage was measured for other refractory radionuclides such as $^{144}$Ce (2.8%), $^{212}$Cm (3%) or $^{239}$Np (3.2%) [29, 51]. Assuming the same released percentage for $^{236}$U or $^{237}$Np, about 111 kg of $^{236}$U and 0.31 kg of $^{237}$Np would have been dispersed by 6 May, 1986. On the other hand, about 25 kg of $^{137}$Cs (80 PBq) were emitted in gaseous forms in the nearly ten days that followed the accident by the burning graphite and inner heating, which represents 30% of the $^{137}$Cs present in the original fuel [51]. Due to the location of the stricken reactor and the meteorological conditions at the time of the accident, the Mediterranean region was subjected to the relatively strong influence of radioactive contamination resulting from the Chernobyl fallout. In the northern parts of the Mediterranean, $^{137}$Cs inventories were increased by 25-100% depending on the geographical, topographical and meteorological factors. In Monaco, the deposition of $^{137}$Cs by the Chernobyl fallout in May 1986 was 3.1 kBq/m$^2$ (4.2x10$^{12}$ atoms/m$^2$), matching the atmospheric fallout deposition up to 1986, of 3.3±0.6 kBq/m$^2$ [28]. However, the deposition of $^{239}$Pu and other transuranic nuclides was much smaller than that of $^{137}$Cs. From the data collected in 1986 and 1989, it was apparent that the terrestrial contamination associated to Pu and, by extrapolation, to other refractory elements, was essentially limited to the restricted zone (i.e. 30 km around Chernobyl) [29]. Nevertheless, radioactive particles have been found in European countries hundreds of km away from the plant [30], and irradiated uranium has been found in soils even at distances more than 200 km to the north of the Chernobyl reactor [31]. This may indicate a different deposition pattern for the actinides and probably element-dependent, than the expected one in a first instance.

In order to estimate the $^{236}$U Chernobyl fallout at the DYFAMED station, two different assumptions have been considered. On one hand, it is assumed a similar dispersion pattern for the released uranium and plutonium isotopes. It has been measured a $^{239+240}$Pu/$^{137}$Cs activity ratio at the level of 5x10$^{-5}$ for the Chernobyl fallout in the Mediterranean region [28]. That ratio translates into a $^{239}$Pu/$^{137}$Cs activity ratio of about 2x10$^{-5}$, or an atomic ratio of 0.016 if the $^{240}$Pu/$^{239}$Pu activity ratio for the Chernobyl debris is used (1.57, or an atom ratio or 0.43 [51]). If this ratio is multiplied by the $^{236}$U/$^{239}$Pu atom ratio reported for the nuclear fuel in [51] (8.7), a $^{236}$U/$^{137}$Cs atom ratio of 0.14 is obtained for the Chernobyl fallout at the studied area. Considering the measured $^{137}$Cs Chernobyl deposition in Monaco (3.1 kBq/m$^2$ or 4.24x10$^{12}$ atoms/m$^2$ [28]), it is obtained a $^{236}$U inventory of about 0.25 ng/m$^2$, or 0.2 kg if integrated over the whole western Mediterranean (Table V). For the second assumption, uranium is supposed to be dispersed in the same way as volatile elements such as $^{137}$Cs. In this case, if the $^{236}$U/$^{137}$Cs atom ratio in the Chernobyl debris, of about 2.6 [51], is multiplied by the $^{137}$Cs Chernobyl fallout deposition in Monaco, it is obtained a $^{236}$U deposition of about 4.3 ng/m$^2$ or 3.7 kg for the western basin. Although the real situation might much closer to the first assumption given the refractory nature of uranium, it is illustrative to bear in mind the second number, which gives a maximum value. From this discussion, it seems reasonable to consider an influence of the Chernobyl accident on the $^{236}$U inventories in the Mediterranean region, due to its abundance in the debris. The global fallout inventories of other transuranic radionuclides such as $^{237}$Np or Pu isotopes, with a minor presence in the debris, wouldn´t be affected. The same might not be true for the other potential $^{236}$U sources considered in this discussion. For instance, $^{237}$Np might be present in significant amounts in the effluents produced in a NRP such as Marcoule, as it happens for the releases from other NRP. Further experimental studies are necessary to explain the excess of $^{236}$U at DYFAMED station.

5. Summary and conclusions

For the first time, $^{236}$U has been studied in the Mediterranean Sea. The obtained $^{236}$U deep-water column inventory at DYFAMED station (Fig. 1), of about 12.6 ng/m$^2$, is significantly higher than the expected one for global fallout at similar latitudes, of about 5 ng/m$^2$. Such a discrepancy hasn´t been observed for other anthropogenic radionuclides such as $^{237}$Np or $^{239,240}$Pu studied at the same station, whose results will be reported elsewhere [42]. This result points out to the presence of additional $^{236}$U sources in the western Mediterranean basin. The input of $^{236}$U associated to Saharan dust outbreaks has been also studied. Overall, it represents about 0.1% of the $^{236}$U inventory associated directly to global fallout. The influence of other natural processes (exchange via Straits, rivers run-off) and of potential local or regional $^{236}$U sources (e.g. Chernobyl fallout, nuclear facilities) are qualitatively evaluated in this work, however, any clear conclusion is drawn. More experimental data are necessary to get an insight into the origin of the excess of $^{236}$U in the western Mediterranean basin.

Acknowledgements

This work has been financed through the projects FIS2012-31853 and FIS2015-69673-P, provided by the Spanish Ministry of Economy and Competitiveness. The IAEA is grateful for the support provided to its Environment Laboratories by the Government of the Principality of Monaco.

Table I.- $^{238}$U results for DYFAMED seawater samples ("DYF-sw"): (i) extrapolated $^{238}$U concentrations in the original samples based on AMS results; (ii) salinity data [33]; (iii) estimated $^{238}$U concentration in the original samples based on salinity data using the formulas given in [40, 41] , and (iv) percentage of total uranium present in the so-called "AMS sample" (Fig. 2). 1σ errors are given. SD stands for standard deviation.

| Sample code | Depth (m) | $^{238}$U ( in "AMS sample", µg/L) | Salinity (psu) | $^{238}$U (from salinity data, µg/L) | Fraction of total U present in "AMS sample" (%) |
|---|---|---|---|---|---|
| DYF-sw-1 | 10 | 1.40 ± 0.12 | 38.22 | 3.53 ± 0.16 | 40 ± 4 |
| DYF-sw-2 | 100 | 1.60 ± 0.07 | 38.34 | 3.54 ± 0.16 | 45 ± 3 |
| DYF-sw-3 | 300 | 1.85 ± 0.09 | 38.57 | 3.56 ± 0.16 | 52 ± 4 |
| DYF-sw-4 | 1000 | 1.84 ± 0.09 | 38.51 | 3.56 ± 0.16 | 52 ± 4 |
| DYF-sw-5 | 2350 | 1.82 ± 0.09 | 38.48 | 3.55 ± 0.16 | 51 ± 4 |
| | | | | Mean: | 48 |
| | | | | SD: | 6 |

Table II.- Obtained $^{236,238}$U results for the DYFAMED seawater samples ("DYF-sw"): (i) $^{236}$U/$^{238}$U AMS atom ratios; (ii) estimated $^{238}$U volume concentrations based on salinity data (also shown in Table I); and (iii) calculated $^{236}$U atom concentrations from the former two sets of results, as it is explained in section 2.3. 1σ errors are given.

| Sample code | Depth (m) | $^{236}$U/$^{238}$U (x10$^{-9}$ atom ratio) | $^{238}$U from salinity data (µg/L) | $^{236}$U (x10$^6$ atoms/L) |
|---|---|---|---|---|
| DYF-sw-1 | 10 | 2.09 ± 0.07 | 3.53 ± 0.16 | 18.6 ± 1.0 |
| DYF-sw-2 | 100 | 2.13 ± 0.05 | 3.54 ± 0.16 | 19.1 ± 0.9 |
| DYF-sw-3 | 300 | 1.74 ± 0.10 | 3.56 ± 0.16 | 15.7 ± 1.1 |
| DYF-sw-4 | 1000 | 1.40 ± 0.07 | 3.56 ± 0.16 | 12.6 ± 0.9 |
| DYF-sw-5 | 2350 | 1.52 ± 0.06 | 3.55 ± 0.16 | 13.7 ± 0.8 |

Table III.- Obtained results for atmospheric particles collected in Monaco during four Saharan Dust events ("SD"): (i) obtained $^{236}$U AMS concentrations; (ii) $^{238}$U mass concentrations in the samples determined by ICPMS and published in [36]; (iii) calculated $^{236}$U/$^{238}$U atom ratios from the former two sets of results, as it is explained in section 3.2; and (iv) $^{236}$U/$^{239}$Pu atom ratios. The $^{239}$Pu results, which were also analyzed in the same sample aliquots by AMS at the CNA, have been also reported in [36]. 1σ errors are given. SDM stands for standard deviation of the mean.

| Atmospheric particles | Sampling date | $^{236}$U (x10$^6$ atoms /g) | $^{238}$U (ug/g) [36] | $^{236}$U/$^{238}$U (x10$^{-9}$ atom ratio ) | $^{236}$U/$^{239}$Pu (atom ratio) |
|---|---|---|---|---|---|
| SD-1 | 20/02/2004 | 62.4 ± 5.5 | 2.020 ± 0.030 | 12.2 ± 1.1 | 0.097 ± 0.009 |
| SD-2 | 01/05/2004 | 58.9 ± 5.4 | 2.035 ± 0.027 | 11.4 ± 1.0 | 0.098 ± 0.009 |
| SD-3 | 29/07/2005 | 50.2 ± 4.4 | 2.071 ± 0.030 | 9.6 ± 0.8 | 0.101 ± 0.009 |
| SD-4 | 26/05/2008 | 58.1 ± 5.6 | 2.605 ± 0.041 | 8.8 ± 0.9 | 0.138 ± 0.014 |
| | Average: | 57.4 | 2.18 | 10.5 | 0.108 |
| | SDM: | 1.1 | 0.06 | 0.3 | 0.004 |

Table IV.- Obtained results for suspended particles at DYFAMED station ("DYF-sp") collected at 200 and 1000 m depth: (i) $^{236}$U AMS concentrations; (ii) $^{238}$U mass concentrations in the samples determined by ICPMS and published in [36]; (iii) calculated $^{236}$U/$^{238}$U atom ratios from the former two sets of results, as it is explained in section 3.2; and (iv) estimated $^{236}$U/$^{239}$Pu atom ratios based on the $^{239}$Pu results also reported in [36]. 1σ errors are given. SDM stands for standard deviation of the mean.

| Sediment traps | Sampling-date start | Sampling-date end | $^{236}$U (x10$^6$ atoms /g) | $^{238}$U (µg/g) [36] | $^{236}$U/$^{238}$U (x10$^{-9}$ atom ratio) | $^{236}$U/$^{239}$Pu (atom ratio) |
|---|---|---|---|---|---|---|
| DYF-sp-1_200 | 04/01/2004 | 18/01/2004 | 5.8 ± 1.5 | 1.76 ± 0.03 | 1.30 ± 0.33 | 0.008 ± 0.002 |
| DYF-sp-2_1000 | | | 20.7 ± 2.5 | 1.78 ± 0.03 | 4.59 ± 0.56 | 0.028 ± 0.003 |
| DYF-sp-3_200 | 15/02/2004 | 29/02/2004 | 27.3 ± 2.9 | 1.64 ± 0.02 | 6.58 ± 0.70 | 0.038 ± 0.004 |
| DYF-sp-4_1000 | | | 27.8 ± 5.4 | 1.70 ± 0.02 | 6.46 ± 1.26 | 0.070 ± 0.014 |
| DYF-sp-5_200 | 29/02/2004 | 14/03/2004 | 27.8 ± 2.9 | 1.71 ± 0.03 | 6.41 ± 0.67 | 0.036 ± 0.004 |
| DYF-sp-6_1000 | | | 26.4 ± 6.2 | 1.72 ± 0.03 | 6.07 ± 1.42 | 0.038 ± 0.009 |
| Dyf-sp-7_200 | 14/03/2004 | 28/03/2004 | 24.4 ± 2.7 | 1.77 ± 0.03 | 5.44 ± 0.60 | 0.032 ± 0.004 |
| DYF-sp-8_1000 | | | 16.9 ± 4.6 | 1.75 ± 0.02 | 3.82 ± 1.04 | 0.022 ± 0.006 |

| | | | | | |
|---|---|---|---|---|---|
| DYF-sp-9_200 | 28/3/2004 | 23/05/2004 | 16.4 ± 2.1 | 1.47 ± 0.02 | 4.41 ± 0.56 | 0.013 ± 0.002 |
| DYF-sp-10_1000 | | | 15.1 ± | 1.59 ± 0.03 | 3.75 ± 0.69 | 0.020 ± 0.004 |
| | | Average: | 21 | 1.7 | 4.9 | 0.031 |
| | | SDM: | 1 | 0.2 | 0.2 | 0.002 |

Table V.- Summary of the estimations of $^{236}$U inventories in the western Mediterranean associated to different sources, expressed in terms of surface deposition, in ng/m$^2$, and, after the integration over the surface of the whole western basin (8.5x10$^5$ km$^2$ [33]), in kg. In the case of Chernobyl fallout, two different assumptions have been taken into account, as it is discussed in section 4.3.

| Sources | Description | $^{236}$U inventories in the western Mediterranean basin | |
|---|---|---|---|
| | | kg | ng/m$^2$ |
| Global fallout | "Well-known" sources | 4.25 | 5 |
| Exchange via Straits | | 0.6 | 0.7 |
| Saharan dust events | | 0.007 | 0.008 |
| Rivers run-off | | 0.06 | 0.07 |
| Chernobyl fallout (first assumption) | Other local or regional sources | 0.2 | 0.25 |
| Chernobyl fallout (second assumption) | | 3.7 | 4.3 |
| Marcoule NRP | | Unknown | |
| Other nuclear facilities | | | |
| $^{236}$U total inventory from "well-known" sources | | 5 kg | 5.8 ng/m$^2$ |
| Estimated $^{236}$U inventory from Chernobyl fallout | | <3.7 kg | < 4.3 ng/m$^2$ |
| DYFAMED $^{236}$U deep-water column inventory | | 10.7 kg | 12.6 ng/m$^2$ |

[1] Radioactivity in the marine environment, National Academic of Sciences, Washington DC: National Academy Press 1971.
[2] H. Craig, K.K. Turekian, The GEOSECS program: 1976–1979, Earth and Planetary Science Letters, 49 (1980) 263-265.
[3] H. Craig, K.K. Turekian, The GEOSECS Program: 1973–1976, Earth and Planetary Science Letters, 32 (1976) 217-219.
[4] B.V.T. Volchok H. L., Folsom T. R., Broecker W. S., Schuert, B.G.S. E. A., Oceanic distributions of radionuclides from nuclear explosions. In: Radioactivity in the Marine Environment., Washington DC: National Academy Press 1971, pp. 42-89.
[5] L.L. Vintró, P.I. Mitchell, K.J. Smith, P.J. Kershaw, H.D. Livingston, Chapter 3 Transuranium nuclides in the world's oceans, in: D.L. Hugh (Ed.) Radioactivity in the Environment, Elsevier, 2005, pp. 79-108.
[6] H.D. Livingston, P.P. Povinec, **A millenium perspective on the contribution of global fallout radionuclides to ocean science**, Health Physics, 2002, pp. 656-668.
[7] G. Steinhauser, A. Brandl, T.E. Johnson, Comparison of the Chernobyl and Fukushima nuclear accidents: A review of the environmental impacts, Science of The Total Environment, 470–471 (2014) 800-817.
[8] M. Aoyama, K. Hirose, Radiometric determination of anthropogenic radionuclides in seawater, in: P.P. Pavel (Ed.) Radioactivity in the Environment, Elsevier, 2008, pp. 137-162.
[9] J. La Rosa, L. Gastaud, L. Lagan, S.-H. Lee, I. Levy-Palomo, P.P. Povinec, E. Wyse, Recent developments in the analysis of transuranics (Np, Pu, Am) in seawater, Journal of Radioanalytical and Nuclear Chemistry, 236 (2005) 427-436.
[10] I. Levy, P.P. Povinec, M. Aoyama, K. Hirose, J.A. Sanchez-Cabeza, J.F. Comanducci, J. Gastaud, M. Eriksson, Y. Hamajima, C.S. Kim, K. Komura, I. Osvath, P. Roos, S.A. Yim, Marine anthropogenic radiotracers in the Southern Hemisphere: New sampling and analytical strategies, Progress in Oceanography, 89 (2011) 120-133.
[11] L. Cao, W. Bu, J. Zheng, S. Pan, Z. Wang, S. Uchida, Plutonium determination in seawater by inductively coupled plasma mass spectrometry: A review, Talanta, 151 (2016) 30-41.
[12] M.E. Ketterer, S.C. Szechenyi, Determination of plutonium and other transuranic elements by inductively coupled plasma mass spectrometry: A historical perspective and new frontiers in the environmental sciences, Spectrochimica Acta Part B: Atomic Spectroscopy, 63 (2008) 719-737.
[13] D. Oughton, P. Day, K. Fifield, Plutonium measurement using accelerator mass spectrometry: Methodology and applications, in: A. Kudo (Ed.) Radioactivity in the Environment, Elsevier, 2001, pp. 47-62.
[14] P. Thakur, G.P. Mulholland, Determination of 237Np in environmental and nuclear samples: A review of the analytical method, Applied Radiation and Isotopes, 70 (2012) 1747-1778.
[15] P. Lindahl, P. Roos, E. Holm, H. Dahlgaard, Studies of Np and Pu in the marine environment of Swedish–Danish waters and the North Atlantic Ocean, Journal of Environmental Radioactivity, 82 (2005) 285-301.
[16] A. Sakaguchi, K. Kawai, P. Steier, F. Quinto, K. Mino, J. Tomita, M. Hoshi, N. Whitehead, M. Yamamoto, First results on 236U levels in global fallout, Science of The Total Environment, 407 (2009) 4238-4242.
[17] M. Christl, J. Lachner, C. Vockenhuber, O. Lechtenfeld, I. Stimac, M.R. van der Loeff, H.-A. Synal, A depth profile of uranium-236 in the Atlantic Ocean, Geochimica et Cosmochimica Acta, 77 (2012) 98-107.
[18] S.H. Lee, P.P. Povinec, E. Wyse, M.A.C. Hotchkis, Ultra-low-level determination of 236U in IAEA marine reference materials by ICPMS and AMS, Applied Radiation and Isotopes, 66 (2008) 823-828.

[19] C. Vockenhuber, I. Ahmad, R. Golser, W. Kutschera, V. Liechtenstein, A. Priller, P. Steier, S. Winkler, Accelerator mass spectrometry of heavy long-lived radionuclides, International Journal of Mass Spectrometry, 223–224 (2003) 713-732.

[20] K.M. Wilcken, L.K. Fifield, T.T. Barrows, S.G. Tims, L.G. Gladkis, Nucleogenic 36Cl, 236U and 239Pu in uranium ores, Nuclear Instruments and Methods in Physics Research Section B: Beam Interactions with Materials and Atoms, 266 (2008) 3614-3624.

[21] M. Christl, N. Casacuberta, J. Lachner, S. Maxeiner, C. Vockenhuber, H.-A. Synal, I. Goroncy, J. Herrmann, A. Daraoui, C. Walther, R. Michel, Status of 236U analyses at ETH Zurich and the distribution of 236U and 129I in the North Sea in 2009, Nuclear Instruments and Methods in Physics Research Section B: Beam Interactions with Materials and Atoms, 361 (2015) 510-516.

[22] E. Chamizo, M. Christl, L.K. Fifield, Measurement of $^{236}$U on the 1 MV AMS system at the Centro Nacional de Aceleradores (CNA), Nuclear Instruments and Methods in Physics Research, Section B: Beam Interactions with Materials and Atoms, 358 (2015) 45-51.

[23] G. Scognamiglio, E. Chamizo, J.M. López-Gutiérrez, A.M. Müller, S. Padilla, F.J. Santos, M. López-Lora, C. Vivo-Vilches, M. García-León, Recent developments of the 1 MV AMS facility at the Centro Nacional de Aceleradores, Nuclear Instruments and Methods in Physics Research Section B: Beam Interactions with Materials and Atoms, 375 (2016) 17-25.

[24] A. Sakaguchi, A. Kadokura, P. Steier, Y. Takahashi, K. Shizuma, M. Hoshi, T. Nakakuki, M. Yamamoto, Uranium-236 as a new oceanic tracer: A first depth profile in the Japan Sea and comparison with caesium-137, Earth and Planetary Science Letters, 333–334 (2012) 165-170.

[25] N. Casacuberta, M. Christl, J. Lachner, M.R. van der Loeff, P. Masqué, H.A. Synal, A first transect of 236U in the North Atlantic Ocean, Geochimica et Cosmochimica Acta, 133 (2014) 34-46.

[26] S.R. Winkler, P. Steier, J. Carilli, Bomb fall-out 236U as a global oceanic tracer using an annually resolved coral core, Earth and Planetary Science Letters, 359–360 (2012) 124-130.

[27] E. Chamizo, M. López-Lora, M. Villa, N. Casacuberta, J.M. López-Gutiérrez, M.K. Pham, Analysis of 236U and plutonium isotopes, 239,240Pu, on the 1MV AMS system at the Centro Nacional de Aceleradores, as a potential tool in oceanography, Nuclear Instruments and Methods in Physics Research, Section B: Beam Interactions with Materials and Atoms, (2014).

[28] UNEP/IAEA, Assessment of the State of Pollution of the Mediterranean Sea by Radioactive Substances, MAP Technical Reports Series No. 62, UNEP, Athens, 1992.

[29] Chapter 5 Terrestrial contamination from chernobyl and other nuclear power station accidents and its radionuclide composition, in: A.I. Yu (Ed.) Radioactivity in the Environment, Elsevier, 2002, pp. 149-211.

[30] R. Pöllänen, I. Valkama, H. Toivonen, Transport of radioactive particles from the chernobyl accident, Atmospheric Environment, 31 (1997) 3575-3590.

[31] S.F. Boulyga, K.G. Heumann, Determination of extremely low 236U/238U isotope ratios in environmental samples by sector-field inductively coupled plasma mass spectrometry using high-efficiency sample introduction, Journal of Environmental Radioactivity, 88 (2006) 1-10.

[32] P.R. Danesi, J. Moreno, M. Makarewicz, D. Louvat, Residual radionuclide concentrations and estimated radiation doses at the former French nuclear weapons test sites in Algeria, Applied Radiation and Isotopes, 66 (2008) 1671-1674.

[33] S.-H. Lee, J.J. La Rosa, I. Levy-Palomo, B. Oregioni, M.K. Pham, P.P. Povinec, E. Wyse, Recent inputs and budgets of 90Sr, 137Cs, 239,240Pu and 241Am in the northwest Mediterranean Sea, Deep Sea Research Part II: Topical Studies in Oceanography, 50 (2003) 2817-2834.

[34] O. Masson, D. Piga, R. Gurriaran, D. D'Amico, Impact of an exceptional Saharan dust outbreak in France: PM10 and artificial radionuclides concentrations in air and in dust deposit, Atmospheric Environment, 44 (2010) 2478-2486.

[35] J. Martín, J.A. Sanchez-Cabeza, M. Eriksson, I. Levy, J.C. Miquel, Recent accumulation of trace metals in sediments at the DYFAMED site (Northwestern Mediterranean Sea), Marine Pollution Bulletin, 59 (2009) 146-153.

[36] M.K. Pham, E. Chamizo, J.L. Mas, J.C. Miquel, J. Martín, I. Osvath, P.P. Povinec, Impact of Saharan dust events on radionuclide levels in Monaco air and in the water column of the northwest Mediterranean Sea, Journal of Environmental Radioactivity, 2016.

[37] J. La Rosa, L. Gastaud, L. Lagan, S.-H. Lee, I. Levy-Palomo, P.P. Povinec, E. Wyse, Recent developments in the analysis of transuranics (Np, Pu, Am) in seawater, Journal of Radioanalytical and Nuclear Chemistry, 236 (2005) 9.

[38] E.P. Horwitz, M.L. Dietz, R. Chiarizia, H. Diamond, A.M. Essling, D. Graczyk, Separation and preconcentration of uranium from acidic media by extraction chromatography, Analytica Chimica Acta, 266 (1992) 25-37.

[39] E.P. Horwitz, M.L. Dietz, R. Chiarizia, H. Diamond, S.L. Maxwell Iii, M.R. Nelson, Separation and preconcentration of actinides by extraction chromatography using a supported liquid anion exchanger: application to the characterization of high-level nuclear waste solutions, Analytica Chimica Acta, 310 (1995) 63-78.

[40] J.M. Pates, G.K.P. Muir, U–salinity relationships in the Mediterranean: Implications for $^{234}$Th:$^{238}$U particle flux studies, Marine Chemistry, 106 (2007) 530-545.

[41] S.A. Owens, K.O. Buesseler, K.W.W. Sims, Re-evaluating the $^{238}$U-salinity relationship in seawater: Implications for the $^{238}$U–$^{234}$Th disequilibrium method, Marine Chemistry, 127 (2011) 31-39.

[42] M. Bressac, I. Levy, E. Chamizo, J.J. La Rosa, J. Gastaud, B. Oregioni, Temporal evolution of the vertical distributon of $^{137}$Cs, $^{239,240}$Pu and $^{237}$Np in the Northwestern Mediterranean Sea, submitted to Limnology & Oceanography, 2016.

[43] J.M. Kelley, L.A. Bond, T.M. Beasley, Global distribution of Pu isotopes and $^{237}$Np, Science of The Total Environment, 237–238 (1999) 483-500.

[44] R. Fukai, Holm, E., Ballestra, S., A note on vertical distribution ofplutonium and americium in the Mediterranean Sea Oceanologica Acta, 1979.

[45] S. Guerzoni, E. Molinaroli, R. Chester, Saharan dust inputs to the western Mediterranean Sea: depositional patterns, geochemistry and sedimentological implications, Deep Sea Research Part II: Topical Studies in Oceanography, 44 (1997) 631-654.

[46] R. Eigl, M. Srncik, P. Steier, G. Wallner, $^{236}$U/$^{238}$U and $^{240}$Pu/$^{239}$Pu isotopic ratios in small (2 L) sea and river water samples, Journal of Environmental Radioactivity, 116 (2013) 54-58.

[47] W. Ludwig, E. Dumont, M. Meybeck, S. Heussner, River discharges of water and nutrients to the Mediterranean and Black Sea: Major drivers for ecosystem changes during past and future decades?, Progress in Oceanography, 80 (2009) 199-217.

[48] L. Borges Silverio, W.d.Q. Lamas, An analysis of development and research on spent nuclear fuel reprocessing, Energy Policy, 39 (2011) 281-289.

[49] F. Eyrolle, S. Charmasson, D. Louvat, Plutonium isotopes in the lower reaches of the River Rhône over the period 1945–2000: fluxes towards the Mediterranean Sea and sedimentary inventories, Journal of Environmental Radioactivity, 74 (2004) 127-138.

[50] F. Quinto, P. Steier, G. Wallner, A. Wallner, M. Srncik, M. Bichler, W. Kutschera, F. Terrasi, A. Petraglia, C. Sabbarese, The first use of $^{236}$U in the general environment and near a shutdown nuclear power plant, Applied Radiation and Isotopes, 67 (2009) 1775-1780.

[51] UNSCEAR, Exposures and effects of the Chernobyl accident, Vol. II, 2000, pp. 115.